\documentclass[a4paper,12pt]{article}

\usepackage{amssymb,amsfonts,amsmath,epsfig,float,graphicx,xcolor}

\usepackage[]{hyperref}

\usepackage{mathrsfs}
\usepackage{parskip}

\title{Duality of quantum geometries}
\author{Jan Naudts\\
Universiteit Antwerpen\\
\small Physics Department, Universiteitsplein 1, 2610 Antwerpen, Belgium\\
\small	\url{Jan.Naudts@uantwerpen.be}\\
\small	\url{https://orcid.org/0000-0002-4646-1190}
}
\date{}

\newcommand{\be}{\begin{equation}}
\newcommand{\ee}{\end{equation}}

\def\Co{{\mathbb C}}

\def\Io{{\mathbb I}}

\def\Mo{{\mathbb M}}

\def\Ro{{\mathbb R}}

\def\Re{\mbox{ Re }}

\newtheorem{theorem}{Theorem}[section]

\newtheorem{proposition}[theorem]{Proposition}

\newtheorem{definition}[theorem]{Definition}
\def\beginproof{\par\strut\vskip 0.1cm\noindent{\bf Proof}\par}

\def\endproof{\par\strut\hfill$\square$\par\vskip 0.5cm}

\newcommand{\upd}{{\rm \,d}}
\def\commut{{\hskip -1.5pt\mbox{\tiny --}}}
\def\red{}

\begin{document}

\maketitle

\begin{abstract}
 Quantum connections are defined by parallel transport operators acting on a Hilbert space.
 They transport tangent operators along paths in parameter space.
 The metric tensor of a Riemannian manifold is replaced by an inner product of
 pairs of operator fields, similar to the inner product of the Kubo-Mori formalism of
 Linear Response Theory. The metric is used to define the dual of a quantum connection.
 The gradient of the parallel transport operators is the quantum vector potential.
 It defines the covariant derivative of operator fields.
 The covariant derivatives are used to quantify the holonomy of the quantum connection.
 It is shown that a quantum connection is holonomic if and only if its dual is holonomic.
 If the parallel transport operators are unitary then an alpha-family of quantum connections
 can be defined in a way similar to Amari's alpha family of connections in Information Geometry.
 The minus alpha connection is the dual of the alpha connection. In particular, the 
 alpha equal zero connection is self-dual.
 An operator field can be combined with a path in parameter space to produce a path in operator
 space. A definition is given for such a path in operator space to be autoparallel.
 The path in parameter space is then a geodesic for the induced
 connection. 
\end{abstract}

\section{Introduction}
 
This research is part of an effort to generalize results of Information Geometry
\cite{AS85,AN00,AS16,AJVS17}
to the non-commutative context. Several approaches have been tried out already.

Chapter 7 of \cite{AN00} starts from a quantum state space formed by density matrices on a Hilbert space. 
Hasegawa \cite{HH93} introduced an alpha family of quantum divergences for pairs of density matrices. 
More generally, the density matrices are replaced by states on a $C^*$-algebra. These states form 
a differentiable manifold that can be studied by the techniques of Information Geometry.
Many authors have followed this approach \cite{NJ18,NJ19,CIJM19,{NJ21a},NJ23,CdNJS24}.
The approach has two drawbacks. Based on works of Araki \cite{AH74} it is argued in \cite{NJ22,NJ23}
that the definition of quantum exponential families is highly non-unique
because of the non-uniqueness of the Radon–Nikodym derivative in a non-commutative context. 
The other drawback is that the approach with a differentiable manifold of quantum states is
far away of the current practices in both Solid State Physics and in Quantum Field Theory.

The present work shifts the emphasis from quantum states to tangent operators.
Intuitively, tangent operators are directional derivatives of smooth paths in the space of bounded
operators on the Hilbert space. This yields a different starting point than when one 
considers paths in the space of quantum states. The latter point of view is worked out in \cite{NJ21}
for a relativistic field of quantum harmonic oscillators. The quantum states are parameterized
by positions in spacetime. The tangent vectors are linear functionals on a von Neumann algebra.
Gauge transformations implement parallel transport along paths in spacetime. Much of this approach
is taken over by the present work. However, spacetime is replaced by an arbitrary finite-dimensional
space of parameters. The usual identification of a vector of parameters with a point in a
differentiable manifold is replaced by an identification with a tangent space of wave functions,
subspace of the Hilbert space. After introduction of an inner product in the way it is done in
the Kubo-Mori formalism \cite{KR57,MH65,FD75,NVW75} this tangent space becomes a
tangent space of operators.

Parameterized statistical models belonging to an exponential family are the central study object of
Information Geometry. Their geometry is described by a connection which is the dual
of the Euclidean connection on the space of canonical parameters. An important result 
(see Theorem 3.6 of \cite{AN00} or Theorem 2 of \cite {NZ24}) states that if a connection is flat
then the dual connection is flat as well. Hence, the geometry of a statistical model
belonging to an exponential family is flat. In the present work the analogue of the 
Euclidean connection is a quantum connection with unitary parallel transport operators.
It is shown in Proposition \ref {inf:prop:duallyflat} below that if a unitary connection
is holonomic then the dual connection is holonomic as well.

 The structure of the paper is as follows.
 Sections \ref {sect:tanop} to \ref {sect:prodform} introduce the notion of a quantum connection
 and of a metric operator.  The dual of a quantum connection is defined in an obvious manner.
 Unitary connections are discussed in Section \ref {sect:dualcon}.
 Starting from Section \ref {sect:vecpot} the parallel transport operators are considered
 as gauge transformations, the directional derivatives of which define the quantum vector potential.
 Sections \ref {sect:forcefield} and \ref {sect:infhol} elucidate the relation that exists between 
 the antisymmetric force tensor and the  holonomy of the quantum connection.
 Sections \ref {sect:embed:auto} and \ref {sect:geo} show how the quantum geometry makes
 the parameter space into a Riemannian manifold.
 
 {\red
 Section \ref {sect:exam} treats the quantum exponential family as an example.
 Next it is shown that the dually flat geometry of Amari \cite{AS85,AN00},
 with its e- and m-connections,
 is a commutative subcase of a dually holonomic quantum geometry. 
 }
 
 Section \ref {sect:discus} contains a discussion of various issues that arise in the previous sections.
 Finally, Section \ref {sect:concl} evaluates the results obtained so far.

 \section{Tangent operators}
 \label{sect:tanop}

In Solid State Physics the Kubo-Mori formalism of Linear Response Theory \cite{KR57,MH65,FD75}
replaces the tangent vectors of a Riemannian manifold by operators
acting on a Hilbert space. For a mathematical foundation see \cite{NVW75,NPV79}.
The Riemannian metric is replaced by a complex inner product/scalar product
defined on pairs of operators.
It is of the form
\be
\label{qg:metric:def}
(X,Y)_\sim=(T\, X\,\Omega,T\,Y\Omega),
\qquad X,Y\in\mathfrak{M}.
\ee
with $X,Y$ bounded operators on $\mathscr H$ belonging to some von Neumann algebra $\mathfrak M$.
The vector $\Omega$ is a normalized element of the Hilbert space $\mathscr H$.
Its is called the wave vector in what follows.
The operator $T$ is a strictly positive self-adjoint operator with domain including $\mathfrak{M}\Omega$.
It is called the metric operator in what follows,

The situation described above can depend on a vector of parameters $\theta_p$, $p=1,2,\cdots,n$,
belonging to some convex open
domain $\Mo$ in $\Ro^n$. For the sake of simplicity the Hilbert space $\mathscr H$ is assumed to be independent
of $\theta$. The von Neumann algebra $\mathfrak{M}$, the metric operator $T$, the wave function $\Omega$
and the inner product receive a lower index $\theta$ to stress there dependence on the parameters $\theta_p$.

In the Literature the inner product $(\cdot,\cdot)_\sim$ is interpreted as a quantum statistical correlation
between pairs $X,Y$ of observable operators. Here, the operators belonging to $\mathfrak{M}$
are called {\em tangent operators} and are considered to be the analogues of the tangent vectors of
a Riemannian manifold.
For convenience, also unbounded operators affiliated with $\mathfrak{M}$ are considered to be tangent operators.
The inner product $(\cdot,\cdot)_\theta$ is the inner product defined in
the tangent space of the quantum manifold at the point labeled $\theta$, i.e.
\be
\label{tan:inner}
(X,Y)_\theta=(T_\theta X\Omega_\theta,T_\theta Y\Omega_\theta),
\qquad X,Y\in\mathfrak{M}_\theta.
\ee

The points of the quantum manifold are the {\em quantum states}.
However, they are not further specified in the present paper.
The quantum manifold is assumed to be fully specified by its tangent spaces.
The conventional identification between the set of parameters $\Mo$ and the manifold is replaced
by an identification between a vector of parameters $\theta$  and the corresponding tangent space.

For convenience, it is assumed throughout the paper that the metric operators $T_\theta$ are bounded operators
with bounded inverse.
This avoids some of the technical discussions about domains of unbounded operators.
In addition, they satisfy 
\[
T_\theta\Omega_\theta=\Omega_\theta,
\qquad\theta\in\Mo.
\]
It is also assumed that $\mathfrak{M}_\theta\Omega_\theta$ is dense in the Hilbert space $\mathscr H$.
This implies that the inner product $(\cdot,\cdot)_\theta$ is non-degenerate.
It makes the linear space $\mathfrak{M}$ into a pre-Hilbert space the completion of which is denoted
$\mathscr H_\theta$.
The adjoint of an operator $X$ on the Hilbert space $\mathscr H$ is denoted $X^\dagger$.
On the other hand, the adjoint of an operator $\hat X$ on  $\mathscr H_\theta$
is denoted $\hat X^*$.

\section{Quantum connections}
\label{sect:qc:dualdef}

\begin{definition}
A {\em quantum connection} $\hat\Pi$ is determined by a collection of 
{\em parallel transport operators} $\Pi(\gamma)^t_s$. 
These are bounded linear operators acting on the Hilbert space $\mathscr H$.
They describe the transport of tangent operators from $\gamma_s$ to $\gamma_t$
along a smooth path $\gamma:\,t\mapsto \gamma_t$ in the space of parameters $\Mo$. 
\end{definition}

The following properties are assumed.
\begin{itemize}
 \item [ A1 ] $\Pi(\gamma)^t_t$ is the identity operator;
 \item [ A2 ] The inverse of $\Pi(\gamma)^t_s$ exists and equals $\Pi(\gamma)^s_t$;
 \item [ A3 ] If $X$ is a tangent operator at $\gamma_s$ then $\Pi(\gamma)^t_s\,X\,\Pi(\gamma)^s_t$
              is a tangent operator at $\gamma_t$, i.e.
            \[\Pi(\gamma)^t_s\,\mathfrak{M}_{s}\,\Pi(\gamma)^s_t=\mathfrak{M}_{t};
            \]
 \item [ A4 ] The wave vector $\Omega_\theta$ appearing in the definition (\ref {qg:metric:def})
 of the quantum metric 
 is covariant in the sense that 
 \[
\Pi(\gamma)^t_s\,\Omega_{s}=[\Pi(\gamma)^s_t]^\dagger\,\Omega_{s}=\Omega_{t}.
 \]
 \end{itemize}
Note that $\mathfrak{M}_{s}$ is written instead of $\mathfrak{M}_{\gamma_s}$ if it is clear which path
$\gamma$ is intended. A similar remark holds for other quantities.

From the transport operators one can derive \cite{KMS51} covariant derivatives and connection coefficients.
This will be done further on.

Introduce operators $\hat\Pi(\gamma)^t_s$ acting on $\mathscr H_s$, defined by
\be
\label{qc:pihatdef}
 \hat\Pi(\gamma)^t_s X=\Pi(\gamma)^t_s\,X\,\Pi(\gamma)^s_t, 
 \qquad X\in\mathfrak{M}_s.
\ee
The operator $\hat\Pi(\gamma)^t_s X$ belongs to $\mathfrak{M}_t$ by assumption A3.

\begin{proposition}
\label{qc:prop:dual}
 Assume that $T_t$ is a bounded operator with bounded inverse.
 Then $\hat\Pi(\gamma)^t_s$ is a bounded operator.
 Its adjoint is given by
 \be
 \label{qc:adj}
 [\hat\Pi(\gamma)^t_s]^*=\hat Z^s_t
 \quad\mbox{ with }
 Z^s_t=T^{-2}_s\,[\Pi(\gamma)^t_s]^\dagger\,T^2_t.
 \ee
\end{proposition}

\beginproof

Calculate
\begin{align*}
 (\hat\Pi(\gamma)^t_s X,\hat\Pi(\gamma)^t_s X)_t
 &=\bigg(T_t\,\Pi(\gamma)^t_s\,X\,\Omega_t,T_t\,\Pi(\gamma)^t_s\,X\,\Omega_t\bigg)\cr 
 &\le ||T_t||^2\,||\Pi(\gamma)^t_s||^2\,||X\,\Omega_t||^2\cr
 &\le ||T_t||^2\,||\Pi(\gamma)^t_s||^2\,||T^{-1}_t||^2\,(X,X)_t.
\end{align*}
This shows that $\hat\Pi(\gamma)^t_s$ is bounded by
\[
 ||\hat\Pi(\gamma)^t_s||_t\le ||T_t||\,||\Pi(\gamma)^t_s||\,||T^{-1}_t||.
\]

Let us finally show (\ref {qc:adj}).
One has
\begin{align*}
[Z^s_t]^{-1}\Omega_s
&=T^{-2}_t\,[\Pi(\gamma)^s_t]^\dagger\,T^2_s\Omega_s\cr
&=Z^t_s\Omega_s\cr
&=\Omega_t.
\end{align*}
Use this to calculate for $X$ in $\mathfrak{M}_s$ and $Y$ in $\mathfrak{M}_t$
\begin{align*}
 (X,\hat Z^s_tY)_s
 &=(T_s\,X\Omega_s,T_s Z^s_t\,Y\,[Z^s_t]^{-1}\,\Omega_s)\cr
 &=(T_s\,X\Omega_s,T_s Z^s_tY\,\Omega_t)\cr
 &=(X\Omega_s,[\Pi(\gamma)^t_s]^\dagger\,T^2_t\,Y\,\Omega_t)\cr
 &=(T_t \,\Pi(\gamma)^t_s\,X\Omega_s,T_t\,Y\,\Omega_t)\cr
 &=(T_t \,\hat \Pi(\gamma)^t_s\,X\Omega_t,T_t\,Y\,\Omega_t)\cr
 &=(\hat \Pi(\gamma)^t_s\,X,T_t\,Y)_t.
\end{align*}
This shows that the adjoint of $\hat \Pi(\gamma)^t_s$ is $\hat Z^s_t$.

\endproof

\section{Dual connections}
\label{sect:dualcon}

In the context of Differential Geometry dual connections are defined relative to
the metric in the tangent space and relative to a path $t\mapsto\gamma_t$ in the manifold. 
In an analoguous way one can define the dual of a quantum connection.

\begin{definition}
\label{qg:dual:def}
A dual quantum connection $\hat \Pi^\star$, if it exists, 
is defined by transport operators $\Pi^*(\gamma)^t_s$
which satisfy
\be
\label{dual:dualdef}
\bigg(\hat\Pi(\gamma)^t_s\,X,\hat\Pi^\star(\gamma)^t_s\,Y\bigg)_{t}
 =\bigg(X,Y\bigg)_{s},
\qquad X,Y\in\mathfrak{M}_{s}.
\ee
\end{definition}

The proof of the following proposition is straightforward.

\begin{proposition}
If the assumptions of Proposition \ref {qc:prop:dual} are satisfied and
the dual quantum connection $\hat \Pi^\star$ exists
then it is unique and it is given by the adjoint of the inverse of $\hat\Pi$, i.e.
\be
\label{dual:expl}
\hat\Pi^\star(\gamma)^t_s=[\hat\Pi(\gamma)^s_t]^*.
\ee
\end{proposition}

The problem of the existence of the dual quantum connection is that a dual defined by (\ref {dual:expl})
should satisfy the assumptions A1, A2, A3, A4 formulated in Section \ref {sect:qc:dualdef}.

\begin{proposition}
\label{dual:prop:exist}
 Assume that the metric operators $T_\theta$ are bounded operators with bounded inverse.
 Assume in addition that $X$ in $\mathfrak{M}_\theta$ implies that $T^{-2}_\theta X^\dagger T^2_\theta$
 belongs to  $\mathfrak{M}_\theta$ as well. 
 Then each quantum connection $\hat\Pi$ has a unique dual w.r.t.~the
 metric operators $T_\theta$. It is given by 
 \be
 \label{dual:dualexplic}
 \Pi^\star(\gamma)^t_s=T^{-2}_t\,[\Pi(\gamma)^s_t]^\dagger\,T^2_s
\ee 
 and it satisfies (\ref {dual:expl}).
\end{proposition}

\beginproof
Let $\Pi^\star(\gamma)^t_s$ be defined by (\ref {dual:dualexplic}),
i.e.
\[
 \Pi^\star(\gamma)^t_s=Z^t_s
 \quad\mbox{ with }\quad Z^t_s=T^{-2}_t\,[\Pi(\gamma)^s_t]^\dagger\,T^2_s.
\]
These operators clearly satisfy assumptions A1, A2 and A4 of Section \ref {sect:qc:dualdef}.

One has for an arbitrary $X$ in $\mathfrak{M}_s$
\begin{align*}
 \hat Z^t_s\,X
 &=Z^t_s\,X\,Z^s_t\cr
 &=T^{-2}_t\,[\Pi(\gamma)^s_t]^\dagger\,T^2_s\, X\, T^{-2}_s\,[\Pi(\gamma)^t_s]^\dagger\,T^2_t\cr
 &=T^{-2}_t\,\bigg(\Pi(\gamma)^t_s\,T^{-2}_s X^\dagger\,T^2_s\,\Pi(\gamma)^s_t \bigg)^\dagger\,T^2_t.
\end{align*}
It is easy to see using twice the additional assumption of the proposition that this operator belongs 
to $\mathfrak{M}_t$.
Hence, assumption A3 is also satisfied.

One concludes that $\hat\Pi^\star$, as defined by (\ref {dual:dualexplic}) satisfies
the assumptions A1 to A4 of Section \ref {sect:qc:dualdef}. Hence, it is a quantum connection.
It then follows from Proposition \ref {qc:prop:dual} that (\ref {dual:expl}) is satisfied.

\endproof

Another way of writing (\ref {dual:dualexplic}) is
\be
T^{2}_s=\Pi^\dagger(\gamma)^t_s\,T^{2}_t\Pi^\star(\gamma)^t_s.
\ee
Hence, if the connection $\hat\Pi$ and its dual $\hat\Pi^\star$ are known and the metric operator {\red $T_{t}$}
is known at a single site $\gamma_t$ in $\Mo$ then it can be calculated all along the path $\gamma$
by the above expression.

\section{Unitary transport}

A special role is played by the quantum connections $\hat\Pi$ for which the parallel transport operators
$\Pi(\gamma)^t_s$ are unitary operators. Such a connection is said to be unitary in what follows.
Unitary connections have an additional symmetry which is exploited
to introduce alpha-families of connections.

From a Physics point of view it is obvious to require that the transport operators
are unitary because then they conserve the normalization of the wave vectors
and conserve the probabilistic interpretation of Quantum Mechanics.

If the connection is unitary and the operators $T_m$ appearing in
the definition (\ref {qg:metric:def}) of the metric are bounded with bounded inverse
then (\ref {dual:dualexplic}) becomes
\be
\label{dual:dualdef3}
 \Pi^\star(\gamma)^t_s=T^{-2}(\gamma_t)\,\Pi(\gamma)^t_s\,T^{2}(\gamma_s).
\ee
This suggests to introduce a family of connections $\hat\Pi_\alpha$ labeled by
a parameter $\alpha$ and defined by the following parallel transport operators
\be
\label{qg:alpha:def}
\Pi_\alpha(\gamma)^t_s=T^{-(1-\alpha)}(\gamma_t)\,\,\Pi_1(\gamma)^t_s\,\,T^{1-\alpha}(\gamma_s),
\ee
where $\hat\Pi_1$ is a unitary connection.
Clearly is $\hat\Pi_{-1}=\hat\Pi_1^\star$.

The folllowing proposition requires a sharpening of the assumptions made in Proposition \ref {dual:prop:exist}.

\begin{proposition}
\label{prop:existalpha}
 Assume that $X\in\mathfrak{M}_\theta$ implies $Y\in\mathfrak{M}_\theta$ for any $Y$
 of the form
 \be
 Y=T^\lambda(\theta)\,X^\dagger\,T^{-\lambda}(\theta)
 \ee
 with $\lambda\in\Ro$.
 Then the parallel transport operators $\Pi_\alpha(\gamma)^t_s$ defined by (\ref {qg:alpha:def})
 satisfy the axioms A1 to A4 of Section \ref {sect:qc:dualdef} and hence determine a quantum connection
 $\hat\Pi_\alpha$.
\end{proposition}

\beginproof

Assumptions A1 and A2 are satisfied in a rather trivial manner.

Take $X$ in $\mathfrak{M}_s$ and calculate using unitarity of the connection
\begin{align*}
 \Pi_\alpha(\gamma)^t_s\,X\,\Pi_\alpha(\gamma)^s_t
 &=
 T^{-(1-\alpha)}(\gamma_t)\,\,\Pi_1(\gamma)^t_s\,\,
 \bigg[T^{1-\alpha}(\gamma_s) \,X\, T^{-(1-\alpha)}(\gamma_s)\bigg]\cr 
 &\quad \times
 \Pi_1(\gamma)^s_t\,\,T^{1-\alpha}(\gamma_t)\cr
 &=
 T^{-(1-\alpha)}(\gamma_t)\,\,\Pi_1(\gamma)^t_s\,\,
 Y^\dagger\,
 \Pi_1(\gamma)^s_t\,\,T^{1-\alpha}(\gamma_t)\cr
 &=T^{-(1-\alpha)}(\gamma_t)\,
 \bigg[\Pi_1(\gamma)^t_s\,Y\Pi_1(\gamma)^s_t\,\bigg]^\dagger\,T^{1-\alpha}(\gamma_t),
\end{align*}
with
\[
 Y=T^{-(1-\alpha)}(\gamma_s)
 \,X^\dagger\,
 T^{(1-\alpha)}(\gamma_s).
\]
Use now the assumption of the proposition to see that $Y$ belongs to $\mathfrak{M}_s$
so that $\Pi_1(\gamma)^t_s\,Y\Pi_1(\gamma)^s_t$ belongs to $\mathfrak{M}_t$.
Apply again the assumption to conclude that the whole expression belongs to $\mathfrak{M}_t$.
This proves that assumption A3 is satisfied.

Finally consider assumption A4.
A calculation gives
\begin{align*}
[\Pi_\alpha(\gamma)^s_t]^\dagger\,\Omega_s
= \Pi_\alpha(\gamma)^t_s\,\Omega_s
 &=T^{-(1-\alpha)}(\gamma_t)\,\,\Pi_1(\gamma)^t_s\,\,T^{1-\alpha}(\gamma_s)\,\Omega_s\cr
 &=\Omega_t.
\end{align*}
This proves A4.

\endproof

A short calculation shows that
\be
\left[\Pi_\alpha(\gamma)^t_s\right]^\dagger=\Pi_{2-\alpha}(\gamma)^s_t.
\ee
Hence,  the connection $\hat\Pi_\alpha$ is unitary 
when $\alpha$ is such that $\Pi_\alpha=\Pi_{2-\alpha}$.
This is of course the case for $\alpha=1$.

An important property is the following.

\begin{proposition}
 The dual connection of $\hat\Pi_\alpha$ is $\hat\Pi_{-\alpha}$.
 In particular, $\hat\Pi_0$ is self-dual.
\end{proposition}

\beginproof

A short calculation using (\ref {dual:dualdef3}) gives
\begin{align*}
 T^2(\gamma_t)\,\Pi^\star_\alpha(\gamma)^t_s 
 &=\Pi^\dagger_\alpha(\gamma)^s_t\,T^2(\gamma_s)\cr
 &=\bigg[T^{-(1-\alpha)}(\gamma_s)\,\Pi(\gamma)^s_t\,T^{1-\alpha}(\gamma_t)\bigg]^\dagger\,T^2(\gamma_s)\cr
 &=T^{1-\alpha}(\gamma_t)\,\Pi^\dagger(\gamma)^s_t\,T^{1+\alpha}(\gamma_s)\cr
 &=T^2(\gamma_t)\,T^{-(1+\alpha)}(\gamma_t)\,\Pi(\gamma)^t_s\,T^{1+\alpha}(\gamma_s)\cr
 &=T^2(\gamma_t)\Pi_{-\alpha}(\gamma)^t_s.
\end{align*}
This implies $\hat\Pi_\alpha^\star=\hat\Pi_{-\alpha}$.

\endproof

\section{Connections of the product form}
\label{sect:prodform}

A subclass of connections is formed by those for which a composition law holds.

A connection $\hat\Pi$ is of the product form if there exists a non-degenerate operator field $V(\theta)$ such that for any smooth path $\gamma$   one has 
   \be
   \label{qg:holo:prod}
           \Pi(\gamma)^t_s=V(\gamma_t)\,V^{-1}(\gamma_s);
   \ee
Such a connection satisfies the composition law
   \be
   \Pi(\gamma)^t_s\,\Pi(\gamma)^s_p=\Pi(\gamma)^t_p 
   \ee
for any path $\gamma$ in $\Mo$.

If a unitary connection $\hat\Pi$ is of the product form then all members
of the corresponding alpha-family are of the product form as well.

\section{The quantum vector potential}
\label{sect:vecpot}

In Differential Geometry connection coefficients $\Gamma^r_{pq}$ can be obtained from the 
parallel transport operators by taking partial derivatives. 
The story here is slightly different due to the appearance of non-commuting operators.

Consider a path in parameter space $\Mo$ of the form
\[
 \gamma_p^\theta:t\mapsto \theta+tg_p
\]
where $g_p$ is the vector in $\Ro^n$ with components $g_p^q$, which equal $1$ if $p=q$ and zero otherwise.
The directional derivative of the parallel transport operator $\Pi$
along the path $\gamma_p^\theta$ at $t=0$ is denoted $A_p(\theta)$.
It is given by
\be
\label{field:qp:def}
A_p(\theta)
=i\hbar\frac{\upd\,}{\upd t}\Pi(\gamma_p^\theta)^t_0 \bigg|_{t=0}
=-i\hbar\frac{\upd\,}{\upd t}\Pi(\gamma_p^\theta)^0_t \bigg|_{t=0}.
\ee
The imaginary number $i$ is added in the definition of $A_p$ to guarantee that the
latter is a Hermitian operator on $\mathscr H$ in the case of a unitary connection.
The constant of Planck $\hbar$ is added for agreement with applications in Physics.

The assumptions needed in what follows are the following.

\begin{itemize}
 \item [ A5 ] The directional derivatives $A_p(\theta)$ exist and are linear so that
for any smooth path $\gamma$ in $\Mo$ one has
\be
\label{field:qvp:lin}
i\hbar\frac{\upd\,}{\upd t}\Pi(\gamma)^t_s\bigg|_{s=t}=\dot\gamma^p\,A_p(\gamma_t);
\ee
 \item [ A6 ] The domains of the operators $A_p(\theta)$, of the commutators
 \[
  \frac{i}{\hbar}[A_p,A_q]_\commut
 \]
 and of the directional derivatives
 \[
 \frac{\partial\,}{\partial \theta^q}A_p(\theta)=\frac{\upd\,}{\upd t}A_p(\theta+tg_q)\bigg|_{t=0}.
\]
 include $\mathfrak{M}_\theta\Omega_\theta$;
 These operators belong to $\mathfrak{M}_\theta$ or are affiliated with it.
\end{itemize}

As a consequence of (\ref {field:qvp:lin}) the field of wave functions $\Omega_\theta$ satisfies
\be
i\hbar\frac{\upd\,}{\upd t}\Omega_t=\dot\gamma^pA_p(\gamma_t)\Omega_t
\ee
along any smooth path $\gamma$ in the parameter space $\Mo$.
This is the Schr\"odinger equation with Hamiltonian
\[
 H_\gamma(t)=\dot\gamma^pA_p(\gamma_t).
\]

\section{Covariant derivatives}

The vector of operators $A_p(\theta)$, $p=1,2,\cdots,n$, is known as the {\em quantum vector potential}.
It is used to define an operator-valued covariant derivative $\nabla_{\dot\gamma}$ by
\be
\label{field:qvp:defcovder}
[\nabla_{\dot\gamma} X]_t=\dot\gamma^p\,[\nabla_p X]_t
\quad\mbox{ with }\quad
[\nabla_p X]_\theta=\frac{\partial X}{\partial \theta^p}+\frac{i}{\hbar}\bigg[A_p(\theta),X(\theta)\bigg]_\commut,
\ee
where $\gamma$ is any smooth path in $\Mo$ and $X$ is an operator-valued field over $\Mo$.

The justification for this definition is that for any smooth path $\gamma$ in the parameter space $\Mo$
and any smooth operator-valued field $X(\theta)$ one has
\be
\frac{\upd\,}{\upd t}\bigg[\Pi(\gamma)^s_t\,X(\gamma_t)\,\Pi(\gamma)^t_s\bigg]\bigg|_{s=t}
=[\nabla_{\dot\gamma} X]_t.
\ee

Defined in this way $\nabla _{\dot\gamma}$ has the properties 
one usually requests a covariant derivative to satisfy:
\begin{itemize}
 \item [1) ] The map 
 \[
n\mapsto \nabla_n X=n^p\nabla_p X  
 \]
 is linear by construction;
 \item [2) ] $\nabla_p X$ is additive in $X$;
 \item [3) ] The product rule holds; Indeed, one has for any function $f(\theta)$
 \begin{align*}
  [\nabla_p fX]_\theta
  &=\frac{\partial fX}{\partial \theta^p}+\frac{i}{\hbar}\bigg[A_p(\theta),f(\theta)\,X(\theta)\bigg]_\commut\cr
  &=\frac{\partial f}{\partial \theta^p}\,X(\theta)+f(\theta)\,\frac{\partial X}{\partial \theta^p}
  +f(\theta)\,\frac{i}{\hbar}\bigg[A_p(\theta),X(\theta)\bigg]_\commut\cr
  &=\frac{\partial f}{\partial \theta^p}\,X(\theta)+f(\theta)\,[\nabla_p X]_\theta.
 \end{align*}

\end{itemize}

The obvious definition of connection coefficients $\Gamma_{qp,r}$ is
\be
\label{qc:concoefdef}
\Gamma_{qp,r}(\theta)
=\Re (\nabla_q A_p,A_r)_\theta,
\ee
where $(\cdot,\cdot)_\theta$ is the inner product defined by (\ref {tan:inner}).
Introduce a metric tensor $g(\theta)$ defined by
\be
\label{covder:gdef}
g_{pq}(\theta)=\Re\bigg[(A_p,A_q)_\theta-(A_p,\Io)_\theta\,(\Io,A_q)_\theta\bigg].
\ee
The matrix $g$ is symmetric and its eigenvalues cannot be negative.

Assume that $g$ is non-degenerate. Then one has
\be
\label{qvp:nabla_exp}
\nabla_q A_p=\Gamma^{r}_{qp}A_r+\cdots,
\ee
where the missing term is orthogonal to the tangent operators $A_p$ 
for the real part of the inner product $(\cdot,\cdot)_\theta$
and where
\[
 \Gamma^{r}_{qp}:=g^{rs}\Gamma_{qp,s}.
\]

\section{The force field tensor}
\label{sect:forcefield}

\begin{figure}
 \caption{Infinitesimal paths considered in Section \ref {sect:forcefield}. }
 \vskip 16pt
 \center\includegraphics[width=6cm]{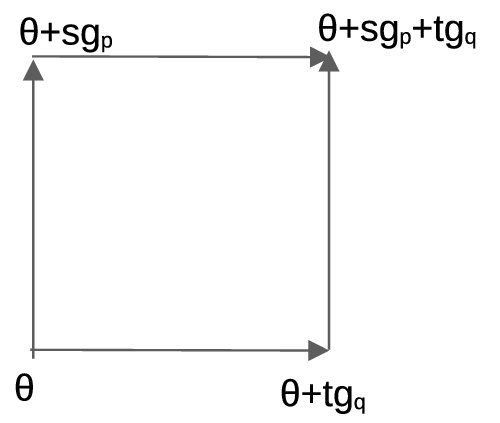}
\end{figure}

Let $\gamma^\theta_p$ denote the path $t\mapsto \theta+tg_p$, as before.
Consider two composed paths which both go from $\theta$ to $\theta+sg_p+tg_q$:
\begin{align}
\label{field:force:twopaths}
 &\theta\rightarrow \theta+sg_p\rightarrow \theta+sg_p+tg_q,\cr
 &\theta\rightarrow \theta+tg_q\rightarrow \theta+sg_p+tg_q.
\end{align}
See the figure.
Introduce operators $H_{pq}(\theta)$ which are formally defined by
\be
\label{qg:H:def}
H_{pq}(\theta)=i\hbar\,\frac{\upd\,}{\upd s}\frac{\upd\,}{\upd t}
\bigg(
\Pi(\gamma^{\theta+tg_q}_p)^s_0\,\Pi(\gamma^\theta_q)^t_0
-\Pi(\gamma^{\theta+sg_p}_q)^t_0\,\Pi(\gamma^\theta_p)^s_0
\bigg)\bigg|_{s=t=0}.
\ee
They are called the {\em holonomy operators} in what follows.

\begin{proposition}
 The holonomy operators satisfy
 \be
\label{field:aft:gap}
H_{pq}(\theta)=F_{pq}(\theta)-\frac{i}{\hbar}\,[A_p(\theta),A_q(\theta)]_\commut,
\ee
with the antisymmetric force field tensor $F$ defined by
\be
\label{field:force:antisymm}
F_{pq}(\theta)=\frac{\partial\,}{\partial \theta^q}A_p(\theta)-\frac{\partial\,}{\partial \theta^p}A_q(\theta).
\ee
\end{proposition}

\beginproof

One has
\begin{align*}
 \frac{\upd\,}{\upd s}\frac{\upd\,}{\upd t}
\left(
\Pi(\gamma^{\theta+tg_q}_p)^s_0\,\Pi(\gamma^\theta_q)^t_0\right)\bigg|_{s=t=0}
&=
-\frac{i}{\hbar}\frac{\upd\,}{\upd t}\,
A_p(\theta+tg_q)\,\Pi(\gamma^\theta_q)^t_0\bigg|_{t=0}\cr
&=-\frac{i}{\hbar} \frac{\upd\,}{\upd t}A_p(\theta+tg_q)\bigg|_{t=0}
-\frac{1}{\hbar^2} A_p(\theta)\,A_q(\theta)\cr
&=
-\frac{i}{\hbar} \frac{\partial\,}{\partial \theta^q}A_p(\theta)-\frac{1}{\hbar^2}\, A_p(\theta)\,A_q(\theta).
\end{align*}
This result implies (\ref {field:aft:gap}).

\endproof

Remember that the partial derivative $\partial A_p(\theta)/\partial \theta^q$
is a shortcut notation for the directional derivative 
\[
 \frac{\partial\,}{\partial \theta^q}A_p(\theta)=\frac{\upd\,}{\upd t}A_p(\theta+tg_q)\bigg|_{t=0}.
\]
In particular, consecutive partial derivatives of operators do not necessarily commute with each other.

If $H_{pq}$ vanishes then one has
\be
F_{pq}=\frac{i}{\hbar}\,[A_p,A_q]_\commut,
\ee
which is a well-known expression for the anti-symmetric force tensor.

Note that (\ref {field:aft:gap}) implies that the operator-valued tensor $H_{pq}$ is anti-symmetric.

Some further properties follow now.

\begin{proposition}
 The covariant derivatives of the elements $A_p$ of the vector potential satisfy
 \be
 \label{field:qvp:nablaAp}
 [\nabla_q A_p]_\theta
 =\frac{\partial\,}{\partial \theta^p}A_q -H_{qp}.
 \ee
\end{proposition}

\beginproof
From the definition (\ref {field:qvp:defcovder}) of $\nabla_q$ one obtains
\[
 [\nabla_q A_p]_\theta=\frac{\partial A_p}{\partial \theta^q}
 +\frac{i}{\hbar}\bigg[A_q(\theta),A_p(\theta)\bigg]_\commut.
\]
Next use (\ref {field:aft:gap}) and (\ref {field:force:antisymm}) to obtain (\ref {field:qvp:nablaAp}).

\endproof

\begin{proposition}
\label{force:prop:prodvan}
 If the connection $\hat\Pi$ is of the product form  (\ref {qg:holo:prod})
 then $H_{pq}$ vanishes.
\end{proposition}

\beginproof
A short calculation gives
\begin{align*}
 H_{pq}(\theta)
 &=i\hbar\,\frac{\upd\,}{\upd s}\frac{\upd\,}{\upd t}
\bigg(
\Pi(\gamma^{\theta+tg_q}_p)^s_0\,\Pi(\gamma^\theta_q)^t_0
-\Pi(\gamma^{\theta+sg_p}_q)^t_0\,\Pi(\gamma^\theta_p)^s_0
\bigg)\bigg|_{s=t=0}\cr
&=i\hbar\,\frac{\upd\,}{\upd s}\frac{\upd\,}{\upd t}
\bigg(
V(\theta+tg_q+sg_p)\,V^{-1}(\theta+tg_q)\,V(\theta+tg_q)\,V^{-1}(\theta)\cr
&\quad
-V(\theta+tg_q+sg_p)\,V^{-1}(\theta+sg_p)\,V(\theta+sg_p)\,V^{-1}(\theta)
\bigg)\bigg|_{s=t=0}\cr
&=0.
\end{align*}

\endproof

\section{Holonomy}
\label{sect:infhol}

The previous section introduces the holonomy operators $H_{pq}$ without elucidating
the connection with holonomy, which is the property that parallel transport
of an operator along a closed loop in parameter space $\Mo$ is an identity operation.
This connection is made here.

\begin{proposition}
\label{infhol:prop:exp}
 If the holonomy operators $H_{pq}(\theta)$ of the previous section exist then the difference
 of parallel transport along the two paths (\ref {field:force:twopaths})
can be expanded as follows
\be
\label{qg:H:exp}
\Pi(\gamma^{\theta+tg_q}_p)^s_0\,\Pi(\gamma^\theta_q)^t_0
-\Pi(\gamma^{\theta+sg_p}_q)^t_0\,\Pi(\gamma^\theta_p)^s_0
 =st\,H_{pq}(\theta)+\mbox{ o}(s)\,+\mbox{ o}(t).
\ee
\end{proposition}

\beginproof
The l.h.s.~vanishes for $s=t=0$. Hence, there is no term of order $s^0t^0$.
The term in $st$ follows from the definition of $H_{pq}$.

If $s=0$ then 
\begin{align*}
\Pi(\gamma^{\theta+tg_q}_p)^s_0\,\Pi(\gamma^\theta_q)^t_0
-\Pi(\gamma^{m_p}_q)^t_0\,\Pi(\gamma^\theta_p)^s_0
 &=
 \Pi(\gamma^\theta_q)^t_0
-\Pi(\gamma^{m}_q)^t_0\cr
&=0.
\end{align*}
Hence, the term in $t$ times $\mbox{ O}(s^0)$.
A similar argument holds for the absence of a term proportional to $s$ times $\mbox{ O}(t^0)$.

\endproof

The expansion (\ref {qg:H:exp}) can be used to calculate the transport along the closed loop 
\[
(\gamma^\theta_p)^0_s\,\circ\,(\gamma_q^{\theta+sg_p})^0_t\,\circ\,
 (\gamma^{\theta+tg_q}_p)^s_0\,\circ\,(\gamma^\theta_q)^t_0.
\]
One finds
\be
\label{infholclosloop}
\Pi(\gamma^\theta_p)^0_s\,\Pi(\gamma_q^{\theta+sg_p})^0_t\,\Pi(\gamma^{\theta+tg_q}_p)^s_0\,\Pi(\gamma^\theta_q)^t_0
=\Io+st H_{pq}(\theta)+\mbox{ o}(s)\,+\mbox{ o}(t).
\ee
Hence, the holonomy operators $H_{pq}$ quantify the  holonomy of the quantum connection $\hat\Pi$.

\section{Curvature}

An obvious definition of an operator-valued curvature tensor $R_{pq}$  is
\[
R_{pq}= \nabla_p\nabla_q-\nabla_q\nabla_p
\]
This commutator of two covariant derivatives is calculated in the following proposition.

\begin{proposition}
\label{field:curv:prop}
 The covariant derivatives $\nabla_p$ and $\nabla_q$ satisfy the commutation relation
 \be
 i\hbar\,(\nabla_p\nabla_q-\nabla_q\nabla_p)\, X=\bigg[H_{pq},X\bigg]_\commut.
 \ee 
\end{proposition}

\beginproof
One calculates
\begin{align*}
[\nabla_p\nabla_q X]_\theta
 &=\frac{\partial\,}{\partial \theta^p}
 \left(\frac{\partial X}{\partial \theta^q}+\frac{i}{\hbar}\left[A_q,X\right]_\commut\right)
 +\frac{i}{\hbar}\bigg[A_p,
 \frac{\partial X}{\partial \theta^q}+\frac{i}{\hbar}\left[A_q,X\right]_\commut\bigg]\cr
 &=
 \frac{\partial^2 X}{\partial \theta^p\partial \theta^q}
  +\frac{i}{\hbar}\bigg[\frac{\partial A_q}{\partial \theta^p},X\bigg]_\commut
  +\frac{i}{\hbar}\bigg[A_q,\frac{\partial X}{\partial \theta^p}\bigg]_\commut\cr
 &+\frac{i}{\hbar}\bigg[A_p,\frac{\partial X}{\partial \theta^q}\bigg]_\commut
  -\frac{1}{\hbar^2}\bigg[A_p,\left[A_q,X\right]_\commut\,\bigg]_\commut.
\end{align*}
This gives
\begin{align*}
[\nabla_p\nabla_q X]_\theta-[\nabla_q\nabla_p X]_\theta
&=\frac{i}{\hbar}\bigg[\frac{\partial A_q}{\partial \theta^p}-\frac{\partial A_p}{\partial \theta^q},X\bigg]_\commut\cr
&-\frac{1}{\hbar^2}\bigg[A_p,\left[A_q,X\right]_\commut\,\bigg]_\commut
 +\frac{1}{\hbar^2}\bigg[A_q,\left[A_p,X\right]_\commut\,\bigg]_\commut\cr
&=
\frac{i}{\hbar}\bigg[F_{qp},X\bigg]_\commut
-\frac{1}{\hbar^2}\bigg[[A_p,A_q]_\commut, X\bigg]_\commut\cr
&=
\frac{i}{\hbar}\bigg[F_{qp},X\bigg]_\commut
+\frac{i}{\hbar}\bigg[F_{pq}-H_{pq},X\bigg]_\commut\cr
&=-\frac{i}{\hbar} \bigg[H_{pq},X\bigg]_\commut.
\end{align*}

\endproof

The proposition shows that the vanishing of the holonomy operators
$H_{pq}$ implies the vanishing of the commutator $[\nabla_p,\nabla_q]_\commut$,
which by definition is the curvature tensor.
Hence, the result shows that holonomy implies absence of curvature.

\section{The dual field operators}

The quantum vector potential of the dual quantum connection $\hat\Pi^\star$ is denoted  $A^\star_p$.
The definition (\ref {field:qp:def}) implies
\be
A^\star_p(\theta)
=i\hbar\frac{\upd\,}{\upd t}\Pi^\star(\gamma_p^\theta)^t_0 \bigg|_{t=0}
=-i\hbar\frac{\upd\,}{\upd t}\Pi^\star(\gamma_p^\theta)^0_t \bigg|_{t=0}.
\ee

\begin{proposition}
\label{dualop:prop:rel}
Assume that the metric operators $T_\theta$ satisfy the conditions of Proposition \ref {dual:prop:exist}.
Then the quantum vector potential $A_p$ and its dual $A^\star_p$ satisfy the following relation
\begin{align}
T_\theta\,A^\star_p(\theta)T^{-1}_\theta-T^{-1}_\theta A^\dagger_p(\theta)\,T_\theta
=-i\hbar\,T^{-1}_\theta\left(\frac{\partial \,}{\partial \theta^p}T^2_\theta\right)\,T^{-1}_\theta.
\label{i:fields:dualAdef}
\end{align}

\end{proposition}

\beginproof
Expression (\ref {dual:dualexplic}) for $\gamma$ equal to $\gamma^\theta_p$ 
with
\[
 \gamma^\theta_p(t)=\theta+tg_p
\]
implies
\[
 T^2_t\,\Pi^\star(\gamma^\theta_p)^t_0=[\Pi(\gamma^\theta_p)^0_t]^\dagger\,T^2_0(\theta).
\]
Take a derivative w.r.t.~$t$ and put $t=0$. This gives
\begin{align*}
 \frac{\partial }{\partial \theta^p} T^2_\theta
 -\frac{i}{\hbar} T^2_\theta\,A^\star_p(\theta)
 &=-\frac{i}{\hbar}A^\dagger_p(\theta)\,T^2_\theta.
\end{align*}
This can be written as (\ref {i:fields:dualAdef}).

\endproof

\begin{proposition}
\label{inf:prop:duallyflat}
 If the connection $\hat\Pi$ is unitary then the holonomy operators $H_{pq}$
 are related to their duals $H^\star_{pq}$ by
 \be
 T_\theta\,H^\star_{pq}\,T^{-1}_\theta=T^{-1}_\theta\,H_{pq}\,T_\theta.
 \ee
 In particular, $H$ vanishes if and only if $H^\star$ vanishes.
\end{proposition}

\beginproof

Use (\ref {qg:H:exp}) and (\ref {dual:dualexplic}) to obtain
\begin{align*}
H^\star_{pq}(\theta)
&=\frac{\upd\,}{\upd s}\frac{\upd\,}{\upd t}
\bigg[\Pi^\star(\gamma^{\theta+tg_q}_p)^s_0\,\Pi^\star(\gamma^\theta_q)^t_0
-\Pi^\star(\gamma^{\theta+sg_p}_q)^t_0\,\Pi^\star(\gamma^\theta_p)^s_0\bigg]\bigg|_{s=t=0}\cr
&=\frac{\upd\,}{\upd s}\frac{\upd\,}{\upd t}
\bigg[
T^{-2}_{\theta+tg_q+sg_p}\cr
&\times
\bigg(
\Pi(\gamma^{\theta+tg_q}_p)^s_0\,\Pi(\gamma^\theta_q)^t_0\,
-
\Pi(\gamma^{\theta+sg_p}_q)^t_0\,\Pi(\gamma^\theta_p)^s_0
\bigg)
\,T^2_\theta\bigg]\bigg|_{s=t=0}\cr
&=
T^{-2}_\theta\,H_{pq}(\theta)\,T^2_\theta
\end{align*}
because if $s=0$ or $t=0$ then
\[
 \bigg(
\Pi(\gamma^{\theta+tg_q}_p)^s_0\,\Pi(\gamma^\theta_q)^t_0\,
-
\Pi(\gamma^{\theta+sg_p}_q)^t_0\,\Pi(\gamma^\theta_p)^s_0
\bigg)
\]
vanishes.

\endproof

\section{Autoparallel operator fields}
\label{sect:embed:auto}

Consider now a path $t\mapsto\gamma_t$ in the parameter space $\Mo$ and an operator field $X(\theta)$.
The induced path $t\mapsto X(\gamma_t)$ satisfies
\be
\label{embed:geo:gen}
\nabla_{\dot\gamma}\,X(\gamma_t)
=\frac{\upd\,}{\upd t}X(\gamma_t)+\frac{i}{\hbar}\dot\gamma^p\,\bigg[A_p(\gamma_t),X(\gamma_t)\bigg]_\commut.
\ee
One can say that the operator field $X$ is autoparallel w.r.t.~the path $\gamma$ if 
the covariant derivative $\nabla_{\dot\gamma}\,X(\gamma_t)$ vanishes.
The physical meaning of this property is
that the quantum measurement of an operator autoparallel w.r.t.~the path $\gamma$
reveals an intrinsic property of a test particle following the path.

More general is the following definition.

\begin{definition}
\label{embed:def:parallel}
 The operator field $X(\theta)$ defined for $\theta$ on the path $t\mapsto\gamma_t\in \Mo$
 is {\em autoparallel} w.r.t.~the path $\gamma$ for the quantum connection $\hat\Pi$
 if it satisfies
 \be
 \label{emb:paralleldef}
 \Pi(\gamma)^t_s\,X(\gamma_s)=X(\gamma_t)\,\Pi(\gamma)^t_s
 \ee
for all $s,t$ in the domain of definition of the path.
\end{definition}

If $X(\theta)$ is defined all over $\Mo$ then one can
take the derivative of (\ref {emb:paralleldef}) w.r.t.~$t$ 
and use (\ref {embed:geo:gen}) to see that the covariant derivative $\nabla_{\dot\gamma}\,X(\gamma_t)$
of an autoparallel operator field $X$ vanishes.

The set of fields $X(\theta)$ autoparallel w.r.t.~the given path $\gamma$ 
forms an algebra. It contains the identity,
which is the constant field $X(\theta)=\Io$. If the connection $\hat\Pi$ is unitary then the adjoint
$X^\dagger(\theta)$ of any element of the algebra does also belong to the algebra.
In that case it is an involutive algebra.

If the connection $\hat\Pi$ is of the product form
with transport operators given by (\ref {qg:holo:prod}) then the operator field $X(\theta)$
belongs to the algebra of autoparallel fields 
if and only if the operator
\[
 V^{-1}(\gamma_t)\,X(\gamma_t)\,V(\gamma_t)
\]
is constant along the path. 
In particular, given a connection $\hat\Pi$ of the product form (\ref {qg:holo:prod}) then
any operator $A$ on $\mathscr H$ defines an autoparallel operator field $X(\theta)$
by
\[
 X(\theta)=V(\theta)\,A\,V^{-1}(\theta).
\]

The proof of the following result is straightforward.

\begin{proposition}
 If the operator field $X(\theta)$ is autoparallel w.r.t.~the path $\gamma$ for the unitary connection
 $\hat\Pi_1$ then
 \[
  T^{-(1-\alpha)}\,X\,T^{1-\alpha}
 \]
 is autoparallel w.r.t.~the path $\gamma$ for the connection  $\hat\Pi_\alpha$.

\end{proposition}

\section{Geodesics}
\label{sect:geo}

The quantum connection $\hat\Pi$ induces a geometry on the parameter space $\Mo$.

\begin{definition}
\label{geo:def:geodef}
 A path $\gamma$ in $\Mo$ is a geodesic for the quantum connection $\hat\Pi$ if 
 the operator field $\dot\gamma^p\, A_p$ is autoparallel w.r.t.~$\gamma$.
\end{definition}

The well-known geodesic equation of Riemannian geometry reads
\be
\label{geo:geoeq}
\ddot\gamma^s+\Gamma^s_{pq}\dot\gamma^p\dot\gamma^q=0.
\ee

\begin{proposition}
 If the path $\gamma$ is a geodesic for the quantum connection $\hat\Pi$ 
 according to Definition \ref {geo:def:geodef} then
 \begin{itemize}
  \item [a) ]
 {\red
 The operator $\dot\gamma^p\,A_p(\gamma_t)$ is constant along the path;}

 \item [b) ]
 {\red
 The quantum vector potential satisfies the equation
 \be
 \label{geo:prop:vp}
 \frac{\upd\,}{\upd t}A_p(\gamma_t)=\frac{i}{\hbar}\bigg[A_p,\dot\gamma^q\,A_q\bigg]_\commut;
 \ee
 } 
  \item [c) ]
 The geodesic equation (\ref {geo:geoeq}) is satisfied
 with connection coefficients $\Gamma^s_{pq}$ given by (\ref {qc:concoefdef}).
 
 \end{itemize}
 
\end{proposition}

\beginproof

\paragraph{a)}
From Definition \ref {embed:def:parallel} of autoparallelism
it follows that if $\gamma$ is a geodesic then one has
\be
 \label{emb:geodef}
 \Pi(\gamma)^t_s\,\dot\gamma^p(s)\,A_p(\gamma_s)
 =\dot\gamma^p(t)\,A_p(\gamma_t)\,\Pi(\gamma)^t_s.
 \ee
Take a derivative of (\ref {emb:geodef}) w.r.t.~$s$ and put $t=s$. Then one obtains
\begin{align*}
 \mbox{lhs}&=\frac{\upd \,}{\upd s}\bigg[\Pi(\gamma)^t_s\,\dot\gamma^p(s)\,A_p(\gamma_s)\bigg]_{s=t}\cr 
 &=\frac{i}{\hbar} \dot\gamma^qA_q\,\dot\gamma^p\,A_p
 +\ddot\gamma^p\,A_p
 +\dot\gamma^p\,\frac{\upd \,}{\upd t}A_p(\gamma_t)
\end{align*}
and
\begin{align*}
 \mbox{rhs}
 &=\frac{\upd \,}{\upd s}\bigg[\dot\gamma^p(t)\,A_p(\gamma_t)\,\Pi(\gamma)^t_s\bigg]_{s=t}\cr
 &=\frac i\hbar \dot\gamma^p\,A_p\,\dot\gamma^q A_q.
\end{align*}
Equating lhs and rhs gives
\begin{align}
 \label{geo:temp}
 {\red
 \frac{\upd\,}{\upd t}\bigg(\dot\gamma^p\,A_p(\gamma_t)\bigg)=
 }
 \ddot\gamma^p\,A_p +\dot\gamma^p\,\frac{\upd \,}{\upd t}A_p(\gamma_t)=0
\end{align}
{\red
This proves the first part of the proposition.
}

\paragraph{b)}
Write using (\ref {field:qvp:nablaAp})
\begin{align}
\label{geo:props:temp}
 \frac{\upd \,}{\upd t}A_p(\gamma_t)
 &=\dot\gamma^q\left[\frac{\partial A_p}{\partial \theta^q}\right]_{\theta=\gamma_t}\cr 
 &=\dot\gamma^q\left[H_{pq}+\left[\nabla_p A_q\right]_{\gamma_t}\right]
\end{align}
{\red
Along the geodesic is 
\[
 \dot\gamma^q\left[\nabla_q A_p\right]_{\gamma_t}=0.
\]
Hence, one has using (\ref {field:qvp:nablaAp})
\begin{align*}
 \dot\gamma^q\left[\nabla_p A_q\right]_{\gamma_t}
 &=\dot\gamma^q\left[\nabla_p A_q-\nabla_q A_p\right]_{\gamma_t}\cr
 &=\dot\gamma^q\left[\frac{\partial A_p}{\partial \theta^q}-\frac{\partial A_q}{\partial \theta^p}
 +2H_{qp}\right]_{\gamma_t}\cr
 &=\dot\gamma^q\left[F_{pq}+2H_{qp}\right].
\end{align*}
In combination with (\ref {geo:props:temp}) this gives 
\[
 \frac{\upd \,}{\upd t}A_p(\gamma_t)=\dot\gamma^q\left[F_{pq}+2H_{qp}\right]
 =\frac{i}{\hbar}\bigg[A_p,\dot\gamma^q\,A_q\bigg]_\commut.
\]

}

\paragraph{c)}
Use the expansion (\ref {qvp:nabla_exp}) to obtain
\[
 \frac{\upd \,}{\upd t}A_p(\gamma_t)=\dot\gamma^q\left[H_{pq}+\Gamma^r_{pq}A_r+\cdots\right],
\]
where the missing term is orthogonal to the operators $A_r$ for the inner product which
is the real part of $(\cdot,\cdot)_t$.
An inner product of (\ref {geo:temp}) with $A_s$ yields
\[
 \bigg(\ddot\gamma^r+\Gamma^r_{pq}\dot\gamma^p\dot\gamma^q\bigg)\,\Re(A_r,A_s)_t=0
\]
where use is made of the anti-symmetry of $H_{pq}$.
By assumption the metric tensor $g$ and {\sl a fortiore} the matrix with entries $\Re(A_r,A_s)_t$
are non-degenerate. Hence, (\ref {geo:geoeq}) follows.

\endproof

{\red
Note that a formal solution of (\ref {geo:prop:vp}) is
\be
\label{geo:formal}
A_p(\gamma_t)
=e^{-i\hbar^{-1}t H(\gamma)}\,A_p(\gamma_0)\,e^{i\hbar^{-1}t H(\gamma)}
\ee
with the constant Hamiltonian $H(\gamma)$ given by
\[
 H(\gamma)=\dot\gamma^p_0\,A_p(\gamma_0).
\]
The proposition shows that if a path $\gamma$ in parameter space
is a geodesic for a
quantum connection $\hat\Pi$ then the quantum vector potential varies
along the geodesic according to (\ref {geo:formal}).
}

\begin{proposition}
\label {geo:prop:constant}
 If $\gamma$ is a geodesic for a self-dual quantum connection $\hat\Pi$ 
 {\red
 and the quantum expectations $(A_p(\gamma_t)\Omega_t,\Omega_t)$ vanish
 }
 then the quantity
 $g_{pq}\dot\gamma^p_t\dot\gamma^q_t$ does not depend on $t$.
\end{proposition}

\beginproof

One has using self-duality followed by (\ref {emb:geodef})
{\red
\begin{align*}
 g_{pq}(\gamma_s)\,\dot\gamma^p_s\dot\gamma^q_s
 &=\Re\bigg(\dot\gamma^p_sA_p(\gamma_s),\dot\gamma^q_sA_q(\gamma_s)\bigg)_s\cr
 &=\Re\bigg(\hat\Pi(\gamma)^t_s\,\dot\gamma^p_sA_p(\gamma_s),
 \hat\Pi(\gamma)^t_s\,\dot\gamma^q_sA_q(\gamma_s)\bigg)_t\cr
 &=\Re\bigg(\dot\gamma^p_tA_p(\gamma_t),\dot\gamma^q_tA_q(\gamma_t)\bigg)_t\cr
 &=g_{pq}(\gamma_t)\,\dot\gamma^p_t\dot\gamma^q_t.
\end{align*}
}
\endproof

{\red
\section{The quantum exponential family}
\label{sect:exam}

\def\Tr{\mbox{ Tr}}
\def\tinyK{{\mbox{\tiny K}}}

A model belonging to the quantum exponential family is 
a parameterized non-degenerate density matrix $\rho_\theta$ of the form
\be
\rho_\theta=\exp(\theta^k E_k-\alpha(\theta)).
\ee
The vector of parameters $\theta$ belongs to some open
convex subset $D$ of $\Ro^n$. The $E_k$ are 
a set of linearly independent $N$-by-$N$ Hermitian matrices.
The normalization function $\alpha(\theta)$ is given by
\be
\alpha(\theta)=\log\Tr \exp(\theta^k E_k).
\ee

A Hermitian matrix $X$ is used to perturb the generator 
$\theta^k E_k$ of the family of density matrices $\rho_\theta$.
The matrix $X$ defines a path $\gamma^X_t$ in the set of density matrices by
\be
\rho_\theta^X(t)=\exp(\theta^k E_k+tX-\alpha(\theta)-\zeta_X(t))
\ee
with 
\[
 \zeta_X(t)=\log\Tr \exp(\theta^k E_k+tX)
-\alpha(\theta).
\]
The tangent to this path at $t=0$ is denoted $\dot\rho^X_\theta$.
It is given by
\[
\dot\rho^X_\theta
=
\frac{\upd\,}{\upd t}\rho_\theta^X(t)\bigg|_{t=0}
=[X]^\tinyK_\theta-\zeta'_X(0)\,\rho_\theta
\quad\mbox{ with }\quad
[X]^\tinyK_\theta=\int_0^1\upd u\,\rho_\theta^u X\rho_\theta^{1-u}.
\]
The quantity $\zeta'_X(0)$ is the derivative of the function $\zeta_X(t)$ at $t=0$.
It satisfies
\[
 \zeta'_X(0)=\Tr\rho_\theta X.
\]

Note that $[X]^\tinyK_\theta$ is known as the Kubo transform of the matrix $X$.
It satisfies 
\begin{align*}
 &[\Io]^\tinyK_\theta=\rho_\theta\qquad
 \mbox{ and } [X^\dagger ]^\tinyK_\theta=\bigg([X]^\tinyK_\theta\bigg)^\dagger\cr
 \mbox{ and } &\Tr [X]^\tinyK_\theta\, Y=\Tr X\,[Y]^\tinyK_\theta\qquad
 \mbox{ and } \Tr [X]^\tinyK_\theta=\Tr\rho_\theta\,X.
\end{align*}

\section{A manifold of density matrices}
\label{sect:densmat}

Consider a parameterized familiy of density matrices $\rho_\theta$,
for instance the quantum exponential family of the previous section.

The representation theorem of Gelfand-Naimark-Segal (GNS)
states that there exists a Hilbert space $\mathscr H$ and 
normalized elements $\Omega_\theta$ in $\mathscr H$ such that
\[
 \Tr\rho_\theta B=(B\Omega_\theta,\Omega_\theta),
 \quad \mbox{ for any  $N$-by-$N$ matrix $B$.}
\]
In addition the space of vectors $B\Omega_\theta$ equals $\mathscr H$.

In what follows an explicit construction of the GNS representation is needed.
Choose the Hilbert space $\mathscr H$ equal to the tensor product
\[
 \mathscr H=\Co^N\otimes\Co^N
\]
of the Hilbert space $\Co^N$ with itself.
Write the density matrix $\rho_\theta$ as
\[
 \rho_\theta=\sum_ip_i(\theta)\,|\psi_i^\theta\rangle\,\langle\psi_i^\theta|
\]
with $(\psi_i^\theta)_i$ a diagonalizing orthonormal basis in $\Co^N$ 
and with $|\psi_i^\theta\rangle\,\langle\psi_i^\theta|$ the orthogonal projection onto 
$\Co\psi_i^\theta$.
Then the wave function $\Omega_\theta$ defined by
\[
 \Omega_\theta=\sum_i\sqrt{p_i(\theta)}\,\psi_i^\theta\otimes\psi_i^\theta
\]
satisfies
\[
\Tr\rho_\theta B=(B\otimes\Io\,\Omega_\theta,\Omega_\theta)
 \quad \mbox{ for any  $N$-by-$N$ matrix $B$.}
\]
See for instance the appendix of \cite{NJ18} for more details about this representation.

Let us now introduce a quantum connection $\hat\Pi(\gamma)$.
Introduce parallel transport operators $\Pi(\gamma)^t_s$
defined by
\[
 \Pi(\gamma)^t_s\psi_i^s\otimes\psi_j^s
 =(p_i(\gamma^t))^{1/2}\,(p_i(\gamma^s))^{-1/2}\,\psi_i^t\otimes\psi_j^t.
\]
They satisfy $\Pi(\gamma)^t_t=\Io$ and $\Pi(\gamma)^t_s\,\Pi(\gamma)^s_t=\Io$ and
$\Pi(\gamma)^t_s\Omega_s=\Omega_t$ and 
 
Let us assume that the space of
tangent operators is maximal, i.e.~of dimension $N^2-1$, both at $\gamma_s$
and at $\gamma_t$ so that tangent operators are mapped onto tangent operators.
Then $\hat\Pi(\gamma)$ satisfies the conditions of Section \ref {sect:qc:dualdef}
for being a quantum connection.

Choose now the metric operator $T_\theta=\rho_\theta^{-1/4}\otimes\Io$.
Then the dual quantum connection $\hat\Pi^\star$ is a unitary connection.
This can be seen as follows. From (\ref {dual:dualexplic}) one obtains
\begin{align*}
\bigg(\Pi^\star(\gamma)^t_s\bigg)^\dagger\,\psi_i^t\otimes\psi_j^t
&=
\rho_s^{-1/2}\,\Pi(\gamma)^s_t\,\rho_t^{1/2}\,\psi_i^t\otimes\psi_j^t\cr
&=
\rho_s^{-1/2}\,\Pi(\gamma)^s_t\,(p_i^t)^{1/2}\,\psi_i^t\otimes\psi_j^t\cr
&=
\rho_s^{-1/2}\,(p_i^s)^{1/2}\,\psi_i^s\otimes\psi_j^s\cr
&=
\psi_i^s\otimes\psi_j^s.
\end{align*}
This shows that $\Pi^\star(\gamma)^t_s$ is the unitary operator which maps
the basis vector $\psi_i^s\otimes\psi_j^s$ onto $\psi_i^t\otimes\psi_j^t$.

Because $\hat\Pi^\star$ is a unitary connection one can apply Proposition \ref {prop:existalpha},
with the roles of $\hat\Pi(\gamma)^t_s$ and $\hat\Pi^\star(\gamma)^t_s$ interchanged, 
to construct an alpha-family of connections.
In particular, the $\alpha=0$ connection $\hat\Pi_0(\gamma)^t_s$ is self-dual.
Its parallel transport operators are given by
\begin{align*}
 \Pi_0(\gamma)^t_s
 &=
 T(\gamma_t)\,\Pi^\star(\gamma)^t_s\,T^{-1}(\gamma_s)
 =T^{-1}(\gamma_t)\,\Pi^\star(\gamma)^t_s\,T(\gamma_s)\cr
 &=\rho_\theta^{1/4}\otimes\Io\,\Pi(\gamma)^t_s\,\rho_\theta^{-1/4}\otimes\Io
\end{align*}
and satisfy
\[
 \Pi_0(\gamma)^t_s\,\psi_i^s\otimes\psi_j^s
 =(p_i^t)^{-1/4}\,(p_i^s)^{1/4}\,\psi_i^t\otimes\psi_j^t.
\]

} 

\section{Discussion}
\label{sect:discus}

\paragraph{A parameterized approach}

Starting point of the paper is the premise that a quantum model is given the states of
which involve a finite number of parameters. This allows to identify the state space with
a subset of the Euclidean space $\Ro^n$. In the study of statistical models belonging
to an exponential family, called Gibbs states in Statistical Physics, this is a common assumption.
Models involving an infinite number of parameters are excluded.
However, an alternative in Information Geometry is the parameter-free approach proposed by
Pistone and Sempi \cite{PS95}. A non-commutative generalization
is studied for instance in \cite{GS01,SRF04a,SRF04b,JA01,JA06}.

\paragraph{The metric}

The inner product often used in the Kubo-Mori formalism is that of
Bogoliubov \cite{PT93}. It describes the statistical correlations between pairs of
operators as observed in equilibrium systems of Quantum Statistical Mechanics.
There is an extensive Literature about such systems and the inner product of Bogoliubov
is an important ingredient of it. See for instance \cite {PD08}.

In applications of the present formalism to quantum field theories other inner products may be
appropriate. 
In the present paper no assumption is made about the existence of quantum states. The manifold $\Mo$
is treated as a space of parameters which label the state of the system whatever this state may be.

\paragraph{Coordinate independence}

A change of coordinates in the space of parameters $\Mo\subset\Ro^n$ results in a change of many of the quantities
studied in the present work. The parallel transport operators $\Pi(\gamma)^t_s$ depend
on the chosen path $\gamma$ in $\Mo$ but not on the choice of basis in $\Ro^n$.
However, the element $A_p$ of the quantum vector potential is a directional derivative in the direction $p$
and hence depends on the choice of basis. It is assumed that the quantum vector potential transforms as a vector
and that the anti-symmetric force tensor $F_{pq}$ and the holonomy operators $H_{pq}$ transform as an 
operator-valued tensor.

\paragraph{Tangent operators}

The definition of tangent operators is kept intentionally vague. In a more concrete context one can define
paths in Hilbert space and in operator space in a more explicit way. See for instance the exponential arcs of
\cite {NJ22,NJ23}. The tangent operators are then directional derivatives along these exponential arcs.

\paragraph{Quantum connections}

The assumptions A1 to A4 made in Section \ref {sect:qc:dualdef} about the properties that
a quantum connection should posses are open for debate. The assumption A3 requires that tangent vectors
are transported onto tangent vectors by the parallel transport operators.
The assumption is non-trivial as is demonstrated by the proof of Proposition \ref {dual:prop:exist}.

Assumption A4 is used at many places.
The parallel transport $\hat\Pi(\gamma)^t_s$ of an operator $X$ from $\gamma_s$ to $\gamma_t$
along a path $\gamma$ is given by (\ref {qc:pihatdef}).
However, when acting on the wave vector $\Omega_s$ one has 
\begin{align*}
 \hat\Pi(\gamma)^t_s X\Omega_s
 &=\Pi(\gamma)^t_s\,X\,\Pi(\gamma)^s_t\Omega_s\cr
 &=\Pi(\gamma)^t_s\,X\,\Omega_t.
\end{align*}
Assumption A4 is essential for this step.
The assumption that
\[
 [\Pi(\gamma)^s_t]^\dagger\,\Omega_{s}=\Omega_{t}
\]
is used in Proposition \ref {dual:prop:exist} to obtain that the adjoint $\hat\Pi^*(\gamma)$ of a
quantum connection $\hat\Pi(\gamma)$ satisfies assumption A4 as well.

\paragraph{Gauge theories}

Parallel transport along a geodesic in parameter space can be considered as a gauge transformation \cite{GD92}.
In principle, one can choose a different Hilbert space $\mathscr{H}$ for each value of the vector
of parameters $\theta$. It is assumed right from the start that all these Hilbert spaces can be identified
with one single copy. This identification of two Hilbert spaces, of the same dimension or separable, is 
unique up to a unitary transformation. This non-uniqueness introduces a gauge freedom in the definition
of the parallel transport operators. Assumptions A3 and A4 partly restrict this gauge freedom.

In the present terminology the connection used in \cite{NJ21} is unitary and of the product form.
In particular, it has vanishing  holonomy operators. This is all right for a model of free-propagating
electromagnetic waves. A more complete description of Quantum Electrodynamics requires
a less-trivial geometry.

\paragraph{The metric tensor}
The subtracted term in the definition (\ref {covder:gdef}) of the metric is omitted in much of the 
Solid State Physics literature because the quantum expectation of the tangent operators 
under consideration vanishes anyhow.

Provost and Vallee \cite{PV80} argue that the subtraction of the product term
is important to obtain invariance of the metric tensor $g$ as given by (\ref {covder:gdef})
under U(1) gauge transformations of the parallel transport operators.
Indeed, a substitution of $\Pi(\gamma)^t_s$ by $\Pi(\gamma)^t_s\,\exp(i(\phi(\gamma_t)-\phi(\gamma_s))$
with $\phi(\theta)$ a real field
implies a substitution of $A_p(\theta)$ by $A_p(\theta)-\hbar\partial\phi/\partial\theta^p$.
The metric tensor $g$ remains invariant because thanks to the subtracted term in (\ref {tan:inner}) 
the additional terms cancel each other out.

\paragraph{Holonomy}

The link between holonomy and Riemannian curvature was studied by Ambrose and Singer \cite{AS53}.
The operators in the l.h.s.~of (\ref {infholclosloop})
belong to what one can call the holonomy group of the quantum connection
at the point $\theta$ of the parameter space $\Mo$.
A study of this group is out of scope of the present paper.
It requires a more detailed setup and other analytic tools.

\paragraph{Operator topologies}

The existence of the directional derivatives in Section \ref {sect:vecpot} and following 
depend on the operator topology one chooses.
If the Hilbert space $\mathscr H$ is finite-dimensional
then one can try to replace these directional derivatives by Fr\'echet derivatives
using the operator norm. 
In the general case Gateaux derivatives of paths in Hilbert space is the best one can hope for.
Then the vector $A_p(\theta)$ in the Hilbert space $\mathscr{H}_\theta$ is  defined by
\begin{align*}
 T_\theta\,A_p(\theta)\,\Omega_\theta
 &=i\hbar \frac{\upd\,}{\upd t}T_\theta\,\Pi(\gamma^\theta_p)^t_0\,\Omega\bigg|_{t=0}\cr
 &=i\hbar \frac{\upd\,}{\upd t}T_\theta\,\Omega_{\theta+tg_p}\bigg|_{t=0}.
\end{align*}
The convergence of such derivatives is out of scope of the present paper and should be studied 
in a more explicit context such as that of \cite{NVW75}.

\paragraph{Berry phase and curvature}

Proposition \ref {infhol:prop:exp} relates the curvature tensor $R_{pq}$ to the holonomy operators $H_{pq}$.
The phenomenon on non-holonomy of the electromagnetic field is known in Solid State Physics
as the Berry phase\cite{BMV84,KSMP17}. It is the phase factor acquired by an eigenvector of the Hamiltonian
when transported along a closed path. This is covered by Proposition \ref {infhol:prop:exp} and its 
consequence (\ref {infholclosloop}).

\paragraph{Geodesics}

The Definition \ref {geo:def:geodef} of a geodesic of a quantum connection 
is rather restrictive because auto-parallelism of vector fields (Definition \ref {embed:def:parallel})
demands equality of operators.
This raises the question whether quantum geodesics exist for arbitrary quantum connections.
Proposition \ref {geo:prop:constant} shows that in the case of a self-dual quantum connection
the length of the tangent vector is constant along the induced geodesic in parameter space.
This suggests that the induced geometry is a metric geometry.
However, in order to prove that, there should exist a geodesic for any choice of initial conditions
in a given point of the parameter space. A proof of this existence is missing.

{\red

\paragraph{The commutative case}

In Section \ref{sect:densmat} a special case ocurs when all density matrices $\rho_\theta$
mutually commute. Then the orthonormal basis $\psi_i^\theta$,
which diagonalizes $\rho_\theta$, can be chosen to be independent of $\theta$.
The dual quantum connection $\hat\Pi^\star(\gamma)$
is then trivial in the sense that
all parallel transport operators $\Pi^\star(\gamma)^t_s$ are equal to the identity
operator $\Io$. This case reproduces the dually flat geometry of Amari \cite{AS85,AN00}
for which it is known that the transport operators of the so-called m-connection
are equal to the identity operator.
One can conclude that in the case of non-commuting density matrices $\rho_\theta$
the present analysis generalizes Amari's dually flat geometry.

}

\section{Conclusion}
\label{sect:concl}

The present work starts from the paradigm that in Quantum Information Geometry
the tangent vectors of a Riemannian manifold are replaced by operator fields
over a parameter space $\Mo$. An inner product for pairs of tangent operators is introduced and
quantum connections are defined by parallel transport operators.
The inner product is used to introduce dual connections.
If the parallel transport operators are unitary then an alpha-family of connections can be defined.
It has the expected property that the minus alpha connection is the dual of the alpha connection.
In particular, the $\alpha=0$ connection is self-dual.
This is the first result of the paper.

The directional derivatives of parallel transport along a path in parameter space
define the quantum vector potential $A_p$. It generalizes the vector potential of
Quantum Electrodynamics. It is used to introduce covariant derivatives and the anti-symmetric
force tensor $F_{pq}$. In a natural way operators $H_{pq}$ show up which add up to the forces $F_{pq}$.
They quantify the  holonomy when transporting the quantum vector potential along a closed path
in parameter space.

The second result of the paper is that if the operators $H_{pq}$ vanish for a quantum connection
then they vanish also for the dual connection. This result mimicks one of the corner
stones of Information Geometry, which is that a connection is flat if and only if its dual is flat.
The analogue of the m-connection of Information Geometry is the unitary quantum connection.
If it has vanishing  holonomy operators then its dual has vanishing  holonomy operators
as well. This dual is the analogue of the e-connection, which characterizes the exponential families
of statistical models.
The vanishing of the  holonomy operators is important because it implies that the two well-known
expressions for the anti-symmetric force tensor coincide.

The final part of the paper considers the geometry of the parameter space $\Mo$
induced by the quantum geometry. It is shown that if the quantum connection is self-dual then the 
induced connection conserves the length of vectors tangent to the geodesic.

{\red
Sections \ref{sect:exam} and \ref {sect:densmat}  show that the concept of
a dually quantum geometry presented in the present work is a genuine
generalization of Amari's dually flat geometry.
}

An application of the present work to Quantum Field Theory in curved spaces is forthcoming.

\section*{}


\begin{thebibliography}{99}


\bibitem{KMS51}
M. S. Knebelman, {\em
Spaces of relative parallelism,}
Ann. Math. {\bf 53}, 387--399 (1951).

\bibitem{AS53}
W. Ambrose, I. Singer, {\em
A theorem on holonomy,}
Trans. Am. Math. Soc. {\bf 75} (3), 428--443 (1953).

\bibitem{KR57}
R. Kubo, {\em
Statistical Mechanical Theory of Irreversible Processes I,}
J. Phys. Soc. Japan, {\bf 12}, 570--586 (1957).

\bibitem{MH65}
H. Mori, {\em
Transport, collective motion, and Brownian motion,}
Progr. Theor. Phys. 33, 423--455 (1965).

\bibitem{AH74}
H. Araki, {\em
Some properties of modular conjugation operator of von Neumann
algebras and a non-commutative Radon–Nikodym theorem with a chain rule,}
Pac. J.  Math. {\bf 50}, 309--354 (1974).

\bibitem{FD75}
D. Forster, {\em
Hydrodynamic fluctuations, broken symmetry, and correlation functions}
(W. A. Benjamin, Inc., 1975)


\bibitem{NVW75}
J. Naudts, A. Verbeure, R. Weder, {\em 
Linear Response Theory and the KMS condition,}
Commun. math. Phys. 44, 87--99 (1975). 


\bibitem{NPV79}
J. Naudts, J. V. Pul{\`e}, A. Verbeure, {\em
Linear response and relaxation in quantum lattice systems,}
J. Stat. Phys. {\bf 21}, 279--288 (1979).

\bibitem{PV80}
J. P. Provost and G. Vallee, {\em
Riemannian structure on manifolds of quantum states,}
Comm. Math. Phys. {\bf 76}, 289--301 (1980).

\bibitem{BMV84}
M. V. Berry, {\em
Quantal Phase Factors Accompanying Adiabatic Changes,}
Proc. Royal Soc. A{\bf 392}, 45--57 (1984).
 
\bibitem{AS85}
S. Amari, {\em
Differential-geometrical methods in statistics, }
Lecture Notes in Statistics {\bf 28}
(Springer, New York, Berlin, 1985)

\bibitem{GD92}
D. J. Gross, {\em
Gauge theory --- Past, Present, and Future?}
Chinese J.  Phys. {\bf 30}, 955--972 (1992).

\bibitem{HH93}
H. Hasegawa, {\em
$\alpha$-divergence of the non-commutative information geometry,}
Rep. Math. Phys. {\bf 33}, 87--93 (1993).

\bibitem{PT93}
D. Petz, G. Toth, {\em
The Bogoliubov inner product in quantum statistics,}
Lett. Math. Phys. {\bf 27}, 205--216 (1993).

\bibitem {PS95}
G. Pistone, C. Sempi, {\em
An infinite-dimensional structure on the space of all 
the probability measures equivalent to a given one,}
Ann. Stat. {\bf 23}, 1543--1561 (1995).

\bibitem{AN00}
S. Amari, H. Nagaoka, {\em
Methods of Information Geometry}
(Oxford University Press, 2000)
(Originally published in Japanese by Iwanami Shoten, Tokyo, Japan, 1993) 

\bibitem{GS01}
M. R. Grasselli and R. F. Streater, {\em 
On the uniqueness of the Chentsov metric in quantum information geometry,}
Infin. Dim. Anal. Quantum Prob. Rel. Top. {\bf 4}, 173--182 (2001).

\bibitem{SRF04a}
R. F. Streater, {\em  
Duality in Quantum Information Geometry,}
Open Syst. \& Inf. Dyn. {\bf 11}, 71--77 (2004).

\bibitem{SRF04b}	
R. F. Streater, {\em  
Quantum Orlicz Spaces in Information Geometry,}
Open Syst. \& Inf. Dyn. {\bf 2004}, {\bf 11}, 359--375 (2004).

\bibitem{JA01}
A. Jen\v cov\'a, {\em 
Geometry of quantum states: Dual connections and divergence functions,}
Rep. Math. Phys. {\bf 47}, 121--138 (2001).

\bibitem{JA06}
A. Jen\v cov\'a , {\em
A construction of a nonparametric quantum information manifold,}
J. Funct. Anal.  {\bf  239}, 1--20 (2006).

\bibitem{PD08}
D. Petz,  {\em
Quantum information theory and quantum statistics}
(Springer, 2008).

\bibitem{AS16}
S. Amari, {\em
Information Geometry and its Applications}
(Springer, 2016)

\bibitem{AJVS17}
N. Ay, J. Jost, H. V{\^an L\^e} and L. Schwachh{\"o}fer, {\em
Information Geometry}
(Springer, 2017)

\bibitem{KSMP17}
M. Kolodrubetz, D. Sels, P. Mehta, A. Polkovnikov, {\em
Geometry and non-adiabatic response in quantum and classical systems,}
Phys. Rep. {\bf 697}, 1--87 (2017).

\bibitem{NJ18}
 J. Naudts, {\em
 Quantum Statistical Manifolds,}
 Entropy {\bf 20}, 472 (2018); correction Entropy {\bf 20}, 796 (2018). 

\bibitem{NJ19}
 J. Naudts, {\em Quantum Statistical Manifolds: The Finite-Dimensional Case,}
 in: Geometric Science of Information, F. Nielsen and F. Barbaresco (Eds.) 
 LNCS 11712 (Springer, 2019), pp. 631 -- 637.

\bibitem{CIJM19}
F. M. Ciaglia, A. Ibort, J. Jost, G. Marmo, {\em 
Manifolds of classical probability distributions
and quantum density operators in infinite dimensions,}
Inf. Geom. {\bf 2}, 231--271 (2019)

\bibitem{NJ21a}
 J. Naudts, {\em Parameter-free description of the manifold of non-degenerate density matrices,}
 Eur. Phys. J. Plus {\bf 136}, 93 (2021). 

\bibitem{NJ21}
J. Naudts, {\em
Gauge transformations of a relativistic field of quantum harmonic oscillators,}
Rep. Math. Phys. {\bf 87}, 15--30 (2021).

\bibitem{NJ22}
J. Naudts, {\em
Exponential arcs in the manifold of vector states on a $\sigma$-finite von {N}eumann algebra,}
Inf. Geom. {\bf 5}, 1--30 (2022).

\bibitem{NJ23}
 J. Naudts, {\em
 Exponential arcs in manifolds of quantum states,}
Front. Phys. {\bf 11}, 1042257, 
(2023).


\bibitem{CdNJS24}
F. M. Ciaglia, F. Di Nocera, J. Jost, L. Schwachh{\"o}fer, {\em
Parametric models and information geometry on
$W^*$-algebras,}
Inf. Geom. {\bf 7}, 329--354 (2024).

\bibitem{NZ24}
J. Naudts and J. Zhang, {\em
Legendre duality: from thermodynamics to information geometry,}
Inf. Geom. {\bf 7},  623--649 (2024).


\end{thebibliography}
\end{document}